\pdfoutput=1
\documentclass[lettersize,journal]{IEEEtran}
\usepackage[table,xcdraw]{xcolor}
\usepackage{amsmath,amsfonts}
\usepackage{algorithmic}
\usepackage{algorithm}
\usepackage{array}
\usepackage[caption=false,font=normalsize,labelfont=sf,textfont=sf]{subfig}
\usepackage{textcomp}
\usepackage{stfloats}
\usepackage{url} 

\usepackage{verbatim}
\usepackage{graphicx}
\usepackage{cite}
\usepackage{slashbox}
\usepackage{multirow}
\usepackage{lscape}
\usepackage{amssymb}
\usepackage{bm}
\usepackage{pifont}
\usepackage{ragged2e}
\usepackage{csquotes}
\usepackage{wrapfig}
\usepackage{mathtools, cuted}
\usepackage{orcidlink}
\usepackage[]{footmisc}
\usepackage{mathptmx}
\usepackage{multicol}

\newcommand{\orcid}[1]{\href{https://orcid.org/#1}{\cellcolor[HTML]{A6CE39}{\aiOrcid}}}

\begin{document}

\title{CAR-BRAINet: Sub-6GHz Aided Spatial Adaptive Beam Prediction with Multi Head Attention for Heterogeneous Vehicular Networks}

\author{Aathira G Menon \textsuperscript{1}\thanks{\noindent\textsuperscript{1}Aathira G Menon, Prabu Krishnan and Shyam Lal are from the Department of Electronics and Communication Engineering, National Institute of Technology Karnataka, Surathkal} \orcidlink{0000-0002-5929-6358} , Prabu Krishnan \textsuperscript{1*} \orcidlink{0000-0001-9730-7997}, Shyam Lal \textsuperscript{1*} \orcidlink{0000-0002-4355-6354}\\
	\thanks{\noindent\textsuperscript{*}prabuk@nitk.edu.in, shyamfec@nitk.edu.in}} 

\maketitle

\begin{abstract}
Heterogeneous Vehicular Networks (HetVNets) play a key role by stacking different communication technologies such as sub-6GHz, mm-wave and DSRC to meet diverse connectivity needs of 5G/B5G vehicular networks. HetVNet helps address the humongous user demands—but maintaining a steady connection in a highly mobile, real-world conditions remain a challenge. Though there has been ample of studies on beam prediction models, a dedicated solution for HetVNets is sparsely explored. Hence, it is the need of the hour to develop a reliable beam prediction solution, specifically for HetVNets. This paper introduces a lightweight deep learning-based solution termed-"CAR-BRAINet" which consists of convolutional neural networks with a powerful multi-head attention (MHA) mechanism. Existing literature on beam prediction is largely studied under a limited, idealised vehicular scenario, often overlooking the real-time complexities and intricacies of vehicular networks. Therefore, this study aims to mimic the complexities of a real-time driving scenario by incorporating key factors such as prominent MAC protocols-3GPP-C-V2X and IEEE 802.11BD, the effect of Doppler shifts under high velocity and varying distance, and SNR levels into three-high quality dynamic datasets pertaining to urban, rural and highway vehicular networks. CAR-BRAINet performs effectively across all the vehicular scenarios, demonstrating precise beam prediction with minimal beam overhead, and a steady improvement of 17.9422\% on the spectral efficiency over the existing methods. Thus, this study justifies the effectiveness of CAR-BRAINet in complex HetVNets, offering promising performance without relying on the location, angle and antenna dimensions of the mobile users, and thereby reducing the redundant sensor-latency.

\end{abstract}


\section{Introduction}
\IEEEPARstart{T}{\lowercase{he}} advancements in 5G and B5G has triggered accelerated demand for ultra-reliable high-speed low-latency connectivity. The Small Cell Networks market is expected to expand from 7.30 million Radio Units (RUs) in 2025 to 8.72 million RUs by 2030 with a compound annual growth rate of 3.61\% \cite{ref01}. Smart city development is a primary driver behind the 5G small cell deployment to achieve superior coverage and enhanced capacity in residential and commercial areas. The placement of macro and small cells at different network levels results in Heterogeneous Vehicular Networks (HetVNets) that provides uninterrupted connectivity and optimised spectrum performance and wider service range across dense regions and secluded areas \cite{ref02_h16}.

\indent The global mobile data traffic is expected to reach 143 exabytes by 2026, as predicted by the Ericsson’s mobility report \cite{ref03_IMP_INTRO_1}. To support this massive user density, 5G has designed Enhanced Mobile Broadband (eMBB). The vast spectrum of the millimeter wave (mm-wave) technology ranging from 30GHz to 300GHz \cite{ref04_IMP_INTRO, ref05_GOOD_INTRO_2, ref06_GOOD_INTRO_3, ref07_GOOD_INTRO_4}, allows high data rate which becomes critical for next-generation networks \cite{ref08_GOOD_INTRO}. But mm-wave suffers from high propagation losses due to harsh environments and obstacles,  \cite{ref09_GOOD_INTRO_5}. Large directional antennas are chosen to provide sufficient beamforming gain, which can stabilise the links between the base station (BST) and mobile user (MU) \cite{ref10_1_VRY_IMP_PLS_REF, ref12_AI_ML}. A high beam-training overhead is associated with such narrow beams. Hence, there is a need to design an efficient beam prediction system (BP) that offers minimal overhead with seamless connectivity in highly mobile heterogeneous vehicular networks (HetVNets) \cite{ref17_IMP_META_1, ref18_IMP_META_2, ref19_IMP_META_3}.

\indent Though there are plenty of research carried out on beam prediction, very few works have addressed beam management (BM) for HetVNets. In contrast, this paper focuses on developing a reliable BP framework to support highly mobile HetVNets by considering significant yet under-recognised characteristics which include user velocity variations, MAC layer data, beam overheads along with driving environments and Doppler effects. These factors remain essential because vehicular networks are dynamic and unpredictable. The geographical and climatic barriers hinders the radio wave propagation in the Indian rural areas\cite{ref20}. Hence this study incorporates numerous experiments across diverse driving scenarios, by integrating these factors to design an intelligent BP model. Our research includes several noteworthy multi-fold contributions such as:
\begin{enumerate}
	\item A regression problem to improve the spectral efficiency of a heterogeneous vehicular network (HetVNet) comprising of the mm-wave, sub-6GHz and dedicated short range communication (DSRC) technology, to achieve a better beam prediction with minimal beam training overhead.
	\item A lightweight deep learning architecture-\textbf{CAR-BRAINet} is developed by incorporating a novel stack of convolutional neural network and multi-head attention to extract the spatial and temporal features effectively to achieve the objective.
	\item A combination of entire sub-6GHz and partial mm-wave/DSRC channel information is used to maximise the spectral efficiency by utilising optimal beam direction predicted by the proposed model.
	\item To mimic the real-time nature of a HetVNets, Urban, Rural and Highway driving scenarios are generated using Wireless InSite, a 3D ray-tracing based software. 
	\item By taking into account of dedicated MAC layer information i.e, C-V2X for mm-wave and IEEE 802.11BD for DSRC technology, geographical blockages and the effect of Doppler shifts caused due to varying velocities, high quality dynamic datasets on diverse driving scenarios are constructed to benchmark the effectiveness of the proposed CAR-BRAINet.
	\item The proposed BP framework is independent of sensor-based informations like location, angle or antenna size of the users, which aids in suppressing delays associated with each of these sensors.

\end{enumerate}

In this study we propose a generalisable attention-driven regression model designed for intelligent and adaptive beam prediction in dynamic vehicular environments termed as CAR-BRAINet (\textbf{\underline{C}}onvolution-\textbf{\underline{A}}ttention \textbf{\underline{R}}egressor for \textbf{\underline{B}}eamforming with \textbf{\underline{R}}easoning \textbf{\underline{A}}nd \textbf{\underline{I}}ntelligent \textbf{\underline{Net}}work). This BP model shall be independent of the location and angle subtended by the mobile users and imposes the least beam training overheads. It will be immune to Doppler shifts and also maintains the lowest possible sensitivity to distance and velocity. The proposed framework shall be trained and evaluated on two prominent MAC protocols.

\indent The following is the structure of the paper. Section \ref{2} discusses the related works. The system model is analysed in Section \ref{3} followed by the proposed model in Section \ref{4}. The model training and experimental setup is detailed in Section\ref{5}. The simulation results are demonstrated in Section \ref{6}, with the paper concluded in Section \ref{7}.

\section{Literature Review}\label{2}

In traditional beam management, exhaustive beam search (EBS) between the BST and MU is used to maximise received power \cite{ref21_28, ref22_IMP_HIER_9, ref23_IMP_HIER_10}. Large antenna arrays and the complex vehicular network imposes a high latency and possible delays in EBS \cite{ref22_IMP_HIER_9, ref23_IMP_HIER_10}. Recent BM approaches delve into non codebook and codebook based techniques \cite{ref08_GOOD_INTRO}. The non-codebook methods is prone to the noise in vehicular scenarios and which degrades the alignment accuracy. On the other hand, codebook based methods estimate the best beam from a pre-defined set and hence entail low overhead and feedback, thereby making them practical as well as deployment friendly.

\indent A beam alignment (BA) method using angle information and channel state information (CSI) of sub-6GHz is proposed in \cite{ref24_15_vry_imp_lr}, while \cite{ref25_7_1_pls_ref} presents a beam training (BT) approach based on MU location for optimal beam selection (BS). Both of these methods rely on traditional computation methods, which struggle in real-time vehicular environments. Zero-forcing precoding-based BS techniques in \cite{ref26_15_1_vry} and \cite{ref27_16_1_vry} depend solely on mm-wave CSI. These non-intelligent methods face high delays due to limited computational efficiency, and accurately estimating mm-wave CSI remains a major challenge in vehicular networks.

In \cite{ref24_15_vry_imp_lr}, a beam alignment (BA) technique based on the angle information and sub-6GHz CSI is proposed, and in \cite{ref25_7_1_pls_ref}, we present a beam training (BT) based on the MU location is discussed. The methodology, however, relies on the use of traditional computation methods which fail in real time vehicular environment. The zero-forcing precoding based BS techniques in \cite{ref26_15_1_vry} and \cite{ref27_16_1_vry} are dependent only on mm-wave CSI. However, due to the lack of computational efficiency, it incurs high delays and poor accuracy in estimating the mm-wave CSI in vehicular networks.

As Deep Learning (DL) effectively handles highly dynamic and complex nature of vehicular networks, it is imperative to employ DL in BM techniques. A precise beam management can be achieved with the aid of DL, as it captures high dimensional features for different driving scenarios \cite{ref04_IMP_INTRO}. Moreover, DL has a better generalisability than traditional mathematical models which are often relied upon their idealised conditions \cite{ref08_GOOD_INTRO}. Other works in this domain utilises mm-wave CSI for beam selection using support vector machine (SVM) \cite{ref28_14_1_vry}, Deep Reinforced Learning (DRL) \cite{ref29_19_1_vry} and Long Short-Term Memory (LSTM) \cite{ref30_18_1_vry, ref31_9_1_vry}. These studies have aimed to minimise the overhead. Nevertheless, it is challenging to extract accurate and noiseless mm-wave CSI in vehicular environments. Hence relying exclusively on  mm-wave based CSI is insufficient while designing a robust BM framework. 

\indent Key features such as the location and velocity of the user, angle of arrival (AoA) and angle of departure (AoD), received signal strength indicator (RSSI), reflection point (RP), power delay profile (PDP) and size of the reflecting arrays are commonly used for beam selection \cite{ref32_23_1_vry, ref33_8_1_vry, ref34_23_17_of_9_1_vry, ref35_12_1_vry, ref36_13_1_vry, ref37_Nearest_1_vry, ref38_10_1_vry, ref39_10_very_imp_lr, ref40_28_very_imp_lr, ref41_27_very_imp_lr, ref42_25_very_imp_lr, ref43_22_1_vry, ref44_30_vry_imp_lr}. Techniques like Convolutional Neural Network (CNN), LSTM, and DRL process these features; however, extracting inputs such as location and angles requires additional sensors and is prone to noise, potentially biasing the model \cite{ref45_Imp_Hierach}. While several studies \cite{ref30_18_1_vry, ref32_23_1_vry, ref33_8_1_vry, ref34_23_17_of_9_1_vry, ref37_Nearest_1_vry, ref42_25_very_imp_lr, ref44_30_vry_imp_lr} address beam overhead reduction, and many overlook it. In regions like India, where geographical obstacles affect signal propagation, incorporating rural scenarios in BM design is crucial \cite{ref20}; yet, no existing studies consider rural settings. Works like \cite{ref47_9_1_vry, ref48_19_1_vry} rely solely on mm-wave CSI, while \cite{ref49_IMP_sub6GHz_Federa} proposes a federated learning-based beamforming (BF) solution using only sub-6GHz CSI, but excludes MAC layer information and validation beyond urban scenarios.

\indent MAC layer information is vital in communication systems as it defines system boundaries \cite{ref50_44_of_18_1_vry_imp}. Incorporating MAC attributes is essential for developing an effective BM solution in a real-time environment. Studies in \cite{ref51_5_1_vry, ref52_6_1_vry} include MAC layers based on IEEE protocols in their BM designs. Given the high mobility in vehicular networks, addressing Doppler shift is critical. However, \cite{ref51_5_1_vry, ref52_6_1_vry} do not validate their models under the effect of Doppler shifts. Meanwhile, \cite{ref53_vry_imp_lr} presents a DDQN-based BA solution considering Doppler effects but lacks evaluation across diverse driving scenarios.\\
\indent HetVNets are crucial for future B5G vehicular applications, making it essential to test BP robustness in heterogeneous networks. While \cite{ref54_het_imp} proposes a BF solution for a HetNet with mm-wave and sub-6GHz, it overlooks key factors like diverse driving scenarios, Doppler effects, and MAC layer information. Limited research addresses the beam management considering these critical aspects.

\indent This paper proposes a novel beam prediction algorithm termed-\textbf{"CAR-BRAINet"} for HetVNets integrating mm-wave, sub-6GHz, and DSRC communication. Due to the difficulty in obtaining perfect mm-wave CSI, we leverage sub-6GHz CSI and partial mm-wave CSI as key features. Given the similarity in channel properties and the practicality of sub-6GHz estimation, this combination forms our primary feature set \cite{ref54_het_imp, ref55_35_of_Het_imp, ref56_36_of_Het_imp}. C-V2X and IEEE 802.11BD, two promising MAC protocols for B5G vehicular networks, are incorporated. To ensure adaptability, diverse driving scenarios—Urban (U), Rural (R), and Highway (H)—are considered. Realistic factors such as geographical obstacles and user density are included. Additionally, this study addresses the Doppler effects associated with high-velocity users. Given the dynamic nature of the vehicular networks, user mobility is also integrated. These factors collectively help generate a large, high-quality dynamic datasets for varied vehicular scenarios, essential to develop and benchmark the proposed DL-based beam prediction model. To the best of our knowledge, no existing works on beam prediction has explored a solution that leverages such a comprehensive set of features, making this study a distinct and novel contribution in the field of V2X communication. Table~\ref{tab:my-table1} presents a comprehensive comparison with prior works.

\begin{table*}[!ht]
\caption{Comparison of existing beam prediction methods against the proposed framework}
\label{tab:my-table1}
\resizebox{\textwidth}{!}{%
	\renewcommand{\arraystretch}{1.5}{%
		\begin{tabular}{ccccccccccccccccccc}
			\hline
			\multicolumn{1}{|c|}{\multirow{2}{*}{\textbf{References}}} & \multicolumn{1}{c|}{\multirow{2}{*}{\textbf{Year}}} & \multicolumn{3}{c|}{\textbf{Features}} & \multicolumn{2}{c|}{\textbf{CSI}} & \multicolumn{3}{c|}{\textbf{Driving scene}} & \multicolumn{2}{c|}{\textbf{MAC}} & \multicolumn{1}{c|}{\multirow{2}{*}{\textbf{HetVNets}}} & \multicolumn{1}{c|}{\multirow{2}{*}{\textbf{\begin{tabular}[c]{@{}c@{}}Doppler\\ effect\end{tabular}}}} & \multicolumn{1}{c|}{\multirow{2}{*}{\textbf{Tool}}} & \multicolumn{1}{c|}{\multirow{2}{*}{\textbf{Dynamic}}} & \multicolumn{1}{c|}{\multirow{2}{*}{\textbf{\begin{tabular}[c]{@{}c@{}}Over\\ head\end{tabular}}}} & \multicolumn{1}{c|}{\multirow{2}{*}{\textbf{\begin{tabular}[c]{@{}c@{}}Model \\ type\end{tabular}}}} & \multicolumn{1}{c|}{\multirow{2}{*}{\textbf{Application}}} \\ \cline{3-12}
			\multicolumn{1}{|c|}{} & \multicolumn{1}{c|}{} & \multicolumn{1}{c|}{\textbf{GPS}} & \multicolumn{1}{c|}{\textbf{Angle}} & \multicolumn{1}{c|}{\textbf{Other}} & \multicolumn{1}{c|}{\textbf{\begin{tabular}[c]{@{}c@{}}mm\\ wave\end{tabular}}} & \multicolumn{1}{c|}{\textbf{\begin{tabular}[c]{@{}c@{}}sub\\ 6GHz\end{tabular}}} & \multicolumn{1}{c|}{\textbf{U}} & \multicolumn{1}{c|}{\textbf{R}} & \multicolumn{1}{c|}{\textbf{H}} & \multicolumn{1}{c|}{\textbf{IEEE}} & \multicolumn{1}{c|}{\textbf{3GPP}} & \multicolumn{1}{c|}{} & \multicolumn{1}{c|}{} & \multicolumn{1}{c|}{} & \multicolumn{1}{c|}{} & \multicolumn{1}{c|}{} & \multicolumn{1}{c|}{} & \multicolumn{1}{c|}{} \\ \hline
			\cite{ref24_15_vry_imp_lr} & 2016 & {\ding{55}} & {\ding{51}} & {\ding{55}} & {\ding{55}} & {\ding{51}} & {\ding{51}} & {\ding{55}} & {\ding{55}} & {\ding{55}} & {\ding{55}} & {\ding{55}} & {\ding{55}} & MM & {\ding{55}} & {\ding{55}} & Non-ML & BA \\ \hline
			\cite{ref27_16_1_vry} & 2018 & {\ding{55}} & {\ding{55}} & {\ding{55}} & {\ding{51}} & {\ding{55}} & {\ding{55}} & {\ding{55}} & {\ding{55}} & {\ding{55}} & {\ding{55}} & {\ding{55}} & {\ding{55}} & SUMO & {\ding{55}} & {\ding{55}} & ZF & BS \\ \hline
			\cite{ref30_18_1_vry} & 2018 & {\ding{55}} & {\ding{55}} & {\ding{55}} & {\ding{51}} & {\ding{55}} & {\ding{55}} & {\ding{55}} & {\ding{55}} & {\ding{55}} & {\ding{55}} & {\ding{55}} & {\ding{55}} & RT & {\ding{55}} & {\ding{51}} & DRL & BS \\ \hline
			\cite{ref33_8_1_vry} & 2019 & {\ding{51}} & {\ding{51}} & Size, RSSI & {\ding{51}} & {\ding{55}} & {\ding{51}} & {\ding{55}} & {\ding{55}} & {\ding{55}} & {\ding{55}} & {\ding{55}} & {\ding{55}} & Dataset & {\ding{51}} & {\ding{51}} & NN & BA \\ \hline
			\cite{ref47_9_1_vry} & 2020 & {\ding{55}} & {\ding{55}} & {\ding{55}} & {\ding{51}} & {\ding{55}} & {\ding{51}} & {\ding{55}} & {\ding{55}} & {\ding{55}} & {\ding{55}} & {\ding{55}} & {\ding{55}} & Dataset & {\ding{55}} & {\ding{51}} & LSTM & BA \\ \hline
			\cite{ref34_23_17_of_9_1_vry} & 2020 & {\ding{55}} & {\ding{55}} & PDP & {\ding{51}} & {\ding{55}} & {\ding{51}} & {\ding{55}} & {\ding{55}} & {\ding{55}} & {\ding{55}} & {\ding{55}} & {\ding{55}} & DeepMIMO & {\ding{55}} & {\ding{51}} & LSTM & BS \\ \hline
			\cite{ref51_5_1_vry} & 2021 & {\ding{51}} & {\ding{55}} & {\ding{55}} & {\ding{55}} & {\ding{55}} & {\ding{51}} & {\ding{55}} & {\ding{51}} & {\ding{51}} & {\ding{55}} & {\ding{55}} & {\ding{55}} & OMNet++ & {\ding{51}} & {\ding{55}} & SAMBA & BF \\ \hline
			\cite{ref49_IMP_sub6GHz_Federa} & 2021 & {\ding{55}} & {\ding{55}} & {\ding{55}} & {\ding{55}} & {\ding{51}} & {\ding{51}} & {\ding{55}} & {\ding{55}} & {\ding{55}} & {\ding{55}} & {\ding{55}} & {\ding{55}} & DeepMIMO & {\ding{55}} & {\ding{51}} & FL & BF \\ \hline
			\cite{ref40_28_very_imp_lr} & 2021 & {\ding{51}} & {\ding{51}} & RP & {\ding{55}} & {\ding{55}} & {\ding{51}} & {\ding{55}} & {\ding{55}} & {\ding{55}} & {\ding{55}} & {\ding{55}} & {\ding{55}} & MM & {\ding{55}} & {\ding{55}} & EBS & BA \\ \hline
			\cite{ref42_25_very_imp_lr} & 2022 & {\ding{51}} & {\ding{55}} & {\ding{55}} & {\ding{51}} & {\ding{55}} & {\ding{51}} & {\ding{55}} & {\ding{55}} & {\ding{55}} & {\ding{55}} & {\ding{55}} & {\ding{55}} & 3GPP & {\ding{55}} & {\ding{51}} & DRL & BT \\ \hline
			\cite{ref44_30_vry_imp_lr} & 2022 & {\ding{55}} & {\ding{51}} & Velocity & {\ding{55}} & {\ding{55}} & {\ding{51}} & {\ding{55}} & {\ding{55}} & {\ding{55}} & {\ding{55}} & {\ding{55}} & {\ding{55}} & DeepMIMO & {\ding{55}} & {\ding{51}} & ML & BS \\ \hline
			\cite{ref37_Nearest_1_vry} & 2022 & {\ding{51}} & {\ding{55}} & {\ding{55}} & {\ding{55}} & {\ding{55}} & {\ding{51}} & {\ding{55}} & {\ding{55}} & {\ding{55}} & {\ding{55}} & {\ding{55}} & {\ding{55}} & DeepMIMO & {\ding{55}} & {\ding{51}} & LSTM & BS \\ \hline
			\cite{ref41_27_very_imp_lr} & 2023 & {\ding{51}} & {\ding{55}} & {\ding{55}} & {\ding{51}} & {\ding{55}} & {\ding{51}} & {\ding{55}} & {\ding{55}} & {\ding{55}} & {\ding{55}} & {\ding{55}} & {\ding{55}} & Dataset & {\ding{55}} & {\ding{55}} & ML & BS \\ \hline
			\cite{ref53_vry_imp_lr} & 2024 & {\ding{51}} & {\ding{55}} & {\ding{55}} & {\ding{55}} & {\ding{55}} & {\ding{51}} & {\ding{55}} & {\ding{55}} & {\ding{55}} & {\ding{55}} & {\ding{55}} & {\ding{51}} & DeepMIMO & {\ding{51}} & {\ding{51}} & DDQN & BA \\ \hline
			\cite{ref54_het_imp} & 2024 & {\ding{55}} & {\ding{55}} & {\ding{55}} & partial & {\ding{51}} & {\ding{51}} & {\ding{55}} & {\ding{55}} & {\ding{55}} & {\ding{55}} & {\ding{51}} & {\ding{55}} & DeepMIMO & {\ding{55}} & {\ding{51}} & GNN & BF \\ \hline
			\cellcolor[HTML]{FDFF26}\textbf{CAR-BRAINet} & \cellcolor[HTML]{FDFF26}\textbf{2025} & \cellcolor[HTML]{FDFF26}\textbf{{\ding{55}}} & \cellcolor[HTML]{FDFF26}\textbf{{\ding{55}}} & \cellcolor[HTML]{FDFF26}\textbf{{\ding{55}}} & \cellcolor[HTML]{FDFF26}\textbf{partial} & \cellcolor[HTML]{FDFF26}\textbf{{\ding{51}}} & \cellcolor[HTML]{FDFF26}\textbf{{\ding{51}}} & \cellcolor[HTML]{FDFF26}\textbf{{\ding{51}}} & \cellcolor[HTML]{FDFF26}\textbf{{\ding{51}}} & \cellcolor[HTML]{FDFF26}\textbf{{\ding{51}}} & \cellcolor[HTML]{FDFF26}\textbf{{\ding{51}}} & \cellcolor[HTML]{FDFF26}\textbf{{\ding{51}}} & \cellcolor[HTML]{FDFF26}\textbf{{\ding{51}}} & \cellcolor[HTML]{FDFF26}\textbf{DeepMIMO} & \cellcolor[HTML]{FDFF26}\textbf{{\ding{51}}} & \cellcolor[HTML]{FDFF26}\textbf{{\ding{51}}} & \cellcolor[HTML]{FDFF26}\textbf{DL} & \cellcolor[HTML]{FDFF26}\textbf{BP} \\ \hline
			\multicolumn{19}{l}{\multirow{2}{*}{ MM = Mathematical Modeling; RT = Real-time test-bed; NN= Neural Network; DNN = Deep Neural Network; DDPG = Deep deterministic policy gradient; DDQN = Double Deep Q Network; BP = Beam Prediction}}
		\end{tabular}
}}%
\end{table*}

\section{System Model}\label{3}
\subsection{System model}

\begin{figure}[!ht]
\centering 
\includegraphics[width=3.5in]{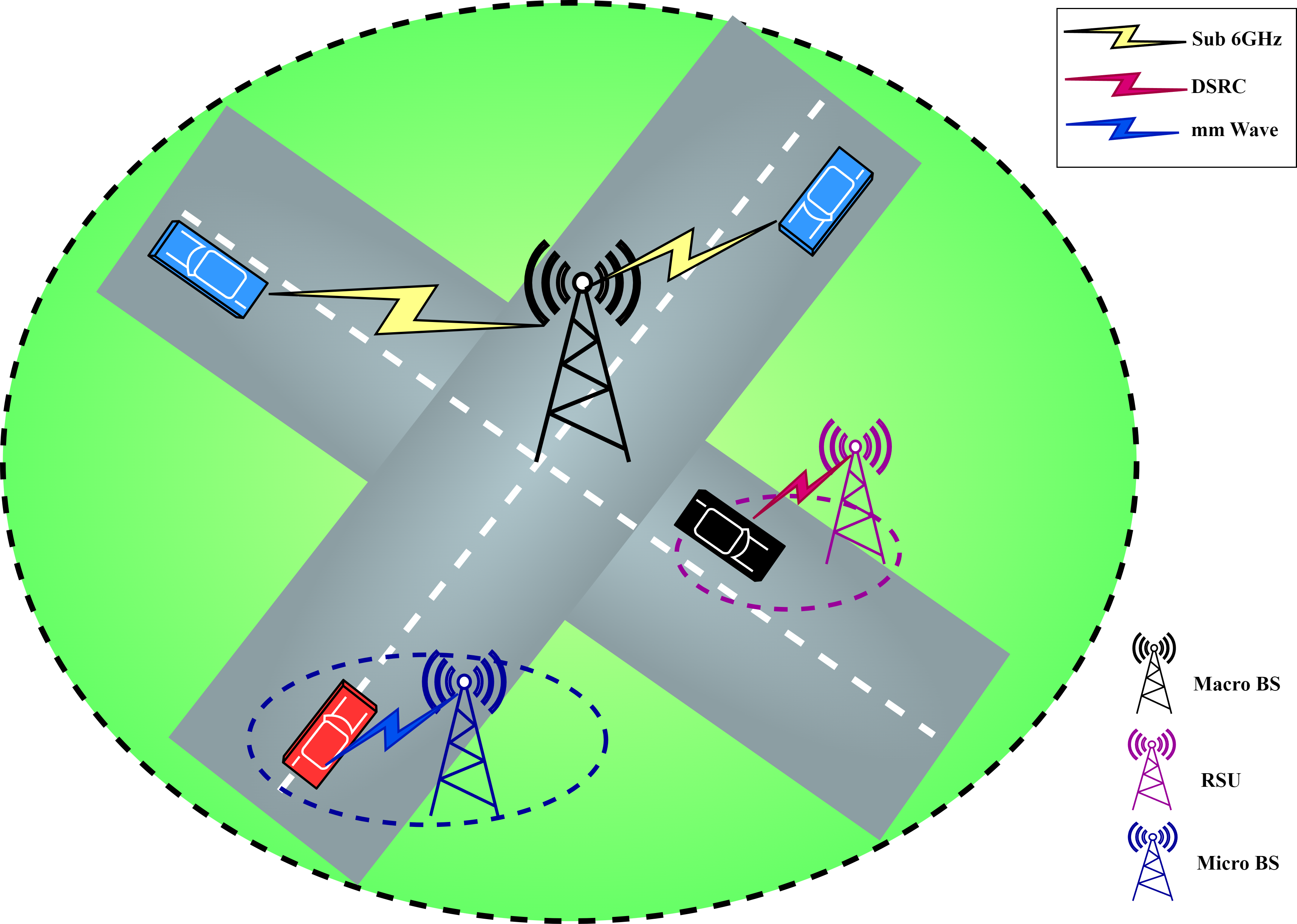}
\caption{An illustration of a Heterogeneous Vehicular Network}
\label{fig:Fig1}
\end{figure}

This study focuses on a heterogeneous vehicular network (HetVNets) integrating DSRC, mm-wave, and sub-6GHz technologies for beam management between the base station (BST) and mobile users (MUs). All users are equipped with receivers supporting all three technologies, where sub-6GHz ensures full coverage, DSRC operates within a smaller coverage area, mainly responsible for transmitting critical safety and control information, and mm-wave provides the highest data rates within a limited area. The study emphasises beam prediction in mm-wave and DSRC, considering diverse driving scenarios, particularly urban, rural and highways (URH), with two prominent technologies, C-V2X and IEEE 802.11BD. A codebook-based beam prediction framework is proposed to handle frequent blockages and challenges in maintaining BST-MU connectivity. Fig.~\ref{fig:Fig1} illustrates a typical scenario representing the system model discussed

Consider $\bm{x_{c,i}}$ as the discrete time transmitted complex base band signal from the \textbf{i\textsuperscript{th}} BST at the \textbf{c\textsubscript{th}} sub-carrier. The received signal \textbf{y\textsubscript{c}} at the sub-carrier c is the transformed in the frequency domain with C-point Fast-Fourier Transform (FFT) and is given in Eq.(\ref{eqn:Eq(1)}).

\begin{equation}
\mathbf{y_{c}} =  \mathrm{\sum_{i=1}^N   \mathbf{h_{c,i} ^{T}  x_{c,i}} +  \mathbf{\mathcal{N}}(0, \sigma ^{2})} 
\label{eqn:Eq(1)}
\end{equation}
Spectral efficiency (SE) is the main performance indicator and is mathematically given by Eq.(\ref{eqn:Eq(3)}). The objective is to maximise SE while balancing latency and beam training overhead.
\begin{equation}
\mathbf{SE} = \mathrm{\frac{1}{C} \sum_{j=1}^C  log_{2}(1+SNR|\sum_{i=1}^N   \mathbf{h_{j,i} ^{T} f_{i} ^{BF}} |^2)}
\label{eqn:Eq(2)}
\end{equation}
where, \textbf{$f_{i} ^{BF}$} represents beam prediction vectors from the codebook
\vspace{-0.2cm}
\subsection{Problem statement}
With reference to a B5G communication network defined above, the problem statement is formulated on maximising the performance metric (SE) while maintaining a promising data rate, minimal latency and beam training overhead. This forms the objective and the set of constraints imposed on the development of the proposed beam prediction pipeline. Mathematically expressed in Eq.(\ref{eqn:Eq(3)}).

\begin{equation}
\mathbf{SE} =   \mathbf{arg max} \mathrm{\sum_{j=1}^C  log_{2}(1+SNR|\sum_{i=1}^N   \mathbf{h_{j,i} ^{T} f_{i} ^{BF}} |^2)}
\label{eqn:Eq(3)}
\end{equation}

\section{Proposed model}\label{4}
Beamforming in a heterogeneous vehicular network faces challenges due to heavy traffic, LOS/NLOS conditions, and rapid variations, requiring intelligent solutions. Traditional complex algorithms add computation overhead and increase latency. To address this, a lightweight and adaptive DL model is designed for efficient and low-latency beam prediction in HetVNets.

\begin{figure*}[!h]
\centering
\includegraphics[width=1.35\columnwidth,height=0.8\columnwidth, keepaspectratio]{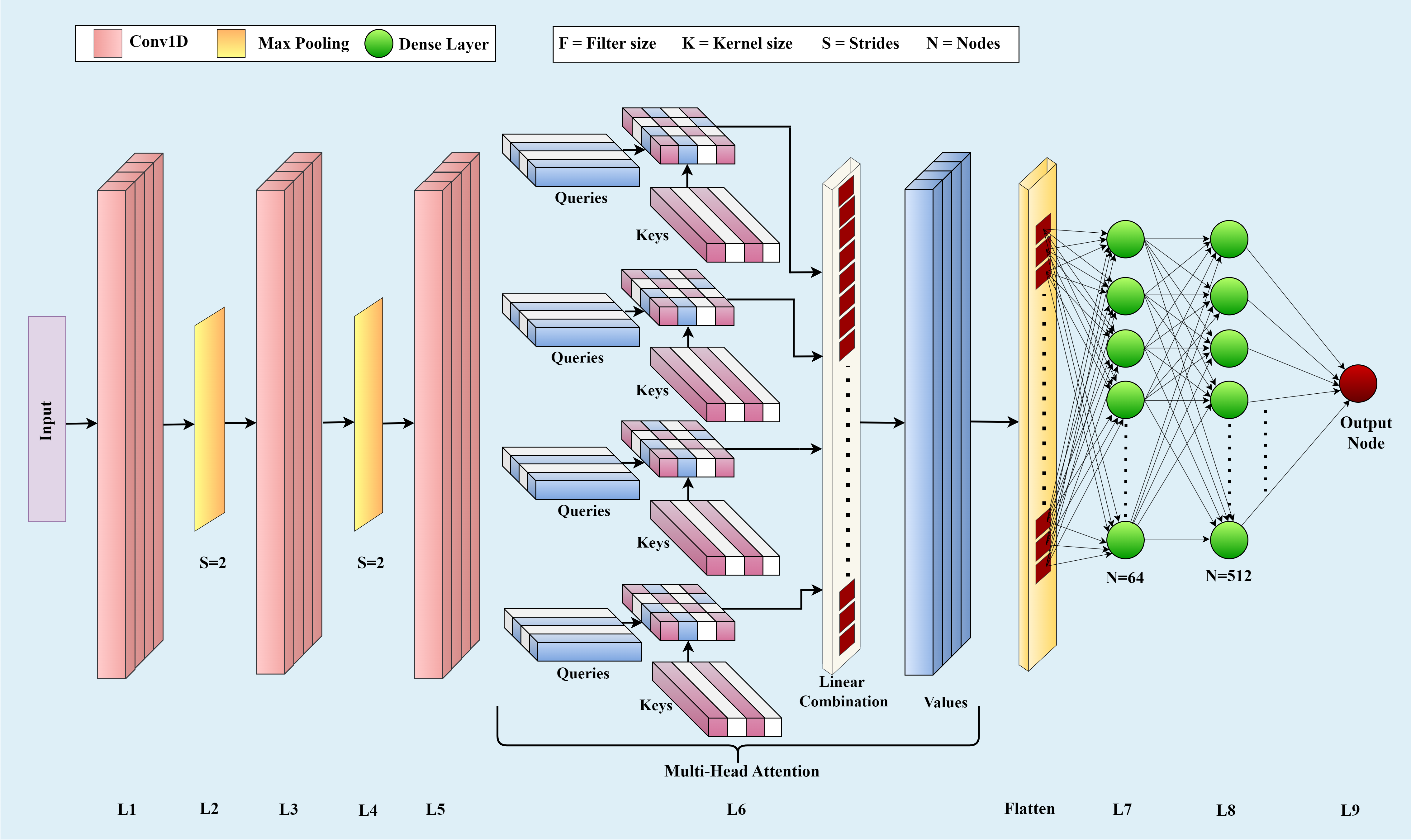}
\caption{Architecture of the proposed CAR-BRAINet}
\label{fig:Fig2}
\end{figure*}

Given the complexity of time-series data, capturing both spatial and temporal dependencies is essential for better performance. To achieve this, a novel CNN-Multi Head Attention (MHA) architecture-\textbf{CAR-BRAINet} is proposed, that effectively captures both feature types. The MHA module enables selective attention to different parts of the input sequence from multiple perspectives, assigning dynamic weights to important segments. Eq.(\ref{eqn:Eq(4)} to Eq.(\ref{eqn:Eq(9)} describes the architectural modelling of CNN with a MHA layer. This attention-driven mechanism helps the model prioritise critical information while suppressing irrelevant parts, ultimately improving its overall performance. Fig.~\ref{fig:Fig2} illustrates the architecture of the proposed model. The pseudo code of the proposed architecture is depicted in Algorithm~\ref{alg:alg1}.

\begin{equation}
\mathrm{F^{(l)}} =  \mathrm{\sigma  \big( W^{(l)} \ast  F^{(l-1)} +  b^{(l)} \big)}
\label{eqn:Eq(4)}
\end{equation}
\vspace{-0.5cm}

\begin{equation}
\mathrm{Q = F W_{Q}} ; \mathrm{K = F W_{K}}; \mathrm{V = F W_{V}}
\label{eqn:Eq(5)}
\end{equation}
\vspace{-0.7cm}

\begin{equation}
\mathrm{Atten(Q, K, V) = softmax \big(\frac{QK^{T}}{\sqrt{d}}\big) V }
\label{eqn:Eq(6)}
\end{equation}
\vspace{-0.7cm}

\begin{equation}
\mathrm{Z = MHA \big(F\big) = Concat( h_{1}, h_{2}, h_{3}, h_{4}) W_{o} }
\label{eqn:Eq(7)}
\end{equation}
\vspace{-0.7cm}

\begin{equation}
\mathrm{D =  \sigma  \big( W_{dense} \ast Z  +  b_{dense}  \big) }
\label{eqn:Eq(8)}
\end{equation}
\vspace{-0.7cm}

\begin{equation}
\mathrm{\hat{y} =  W_{reg}  \ast D  +  b_{reg} }
\label{eqn:Eq(9)}
\end{equation}
\vspace{-0.5cm}

where, \textbf{F} indicates the input feature set; \textbf{W} and \textbf{b} indicate the weight and bias respectively. This study considers MHA with four heads namely \textbf{{h\textsubscript{1}, h\textsubscript{2}, h\textsubscript{3}, h\textsubscript{4}}}. The query, key and value associated with the MHA are indicated by \textbf{Q}, \textbf{K} and \textbf{V}. The regression output is given by $\bm{\hat{y}}$.

\begin{algorithm}[!h]
\caption{CNN-MHA based Regressor-CAR-BRAINet}\label{alg:alg1}
\begin{algorithmic}[1]
	\STATE {\textbf{INPUT:}} Normalised Channel Matrix (\textbf{\underline{\underline{H}}})
	\STATE {\textbf{OUTPUT:}} Spectral Efficiency (\textbf{\underline{SE}})
	\STATE \textbf{\underline{\underline{F}}} = Codebook of Beam-vectors
	\STATE \textbf{begin:}
	\STATE for \textbf{i} = 1, 2, ... \textbf{N\textsubscript{B}}: \hspace{1cm}
	\STATE \hspace{0.5cm}for \textbf{j} in \textbf{e}: \hspace{1cm} 
	\STATE \hspace{1cm} Feed \textbf{\underline{\underline{H}}} into the model for feature-extraction
	\STATE \hspace{1cm} TRAIN(.)			\hspace{1cm}// Model training
	\STATE \hspace{1cm} \bm{{$\hat{y}_j^\theta$}} $\gets$ PRED(.) \hspace{1cm}// Model predictions
	\STATE \hspace{1cm} \bm{$J(\theta)$} $\gets$ $\frac{1}{N}  \sum_{ j =1}^N \bm{L_{\delta}} \big(\hat{y}_j^\theta, y_{j}\big) $
	\STATE \hspace{1cm} optimise the weights using \textbf{Adam} w.r.t $\mathcal{L(.)}$ 
	\STATE \hspace{0.5cm} \textbf{endfor}
	\STATE Save the best prediction $\forall$ \textbf{N\textsubscript{B}}
	\STATE \textbf{endfor}
	\STATE $\textbf{\underline{\underline{$\hat{\bm{H_{eff}}}$}}}$ $\gets$ ${\underline{\underline{\bm{H^{H}}}}}$ $\times$ \underline{\underline{\textbf{F[$\hat{y}$]}}} \\ 
	\STATE \textbf{\underline{SE}} $\gets$ $\sum_1^{N\textsubscript{U}}$ $log_2$(1 + \textbf{SNR} $\times$  $|$\underline{\underline{$\hat{\bm{H_{ eff_{jj}}}}$}}$|$ $^{2}$)
	\STATE \textbf{end}
	\STATE \textbf{N} = size of training data, \bm{$\theta$} = Model parameters,\\ \bm{${L_{ \delta }}}$ = Huber Loss (Eq.(\ref{eqn:Eq(12)})), \textbf{N\textsubscript{B}} = Number of Base-Stations, \textbf{N\textsubscript{U}} = Number of Users, \textbf{e} = Epochs
\end{algorithmic}
\label{alg1}
\end{algorithm}

\section{Training and Implementation}\label{5}
\subsection{Dataset}
Three high-quality datasets representing urban, rural, and highway vehicular scenarios are generated using 3D ray-tracing from the Remcom Wireless InSite simulator-DeepMIMO\cite{ref35}. Scenario-specific network attributes (Table~\ref{tab:my-table2}) are fed into the DeepMIMO generator. User density is set as per the annual report of Telecom Regulatory Authority of India (TRAI) 2023-24 \cite{ref38_3} to emulate near-real-time conditions. MAC layer parameters for C-V2X and IEEE 802.11BD, sourced from~\cite{ref06_GOOD_INTRO_3, ref41, ref42}, are integrated. Focusing on sub-6GHz and partial mm-wave/DSRC channels, the respective channel matrices are stacked, and for the ground truth the spectral efficiency is deduced with the help of the beam direction rendered by the proposed model. The datasets are normalised and made ready for the DL-pipeline.

\begin{table}[!h]
\tiny
\caption{Network Attributes of the Dataset}
\label{tab:my-table2}
\resizebox{0.5\textwidth}{!}{%
	\renewcommand{\arraystretch}{1.15}{%
		\begin{tabular}{l|ccc}
			\hline
			\multicolumn{1}{c|}{\textbf{Data Attributes}}          & \multicolumn{3}{c}{\textbf{Value}}            \\ \hline
			\multirow{3}{*}{\textbf{System Bandwidth}} & Sub-6GHz (C-V2X)                               & \multicolumn{2}{c}{20MHz}                              \\ \cline{2-4} 
			& DSRC-IEEE 802.11BD     & \multicolumn{2}{c}{20MHz} \\
			\cline{2-4} 
			& mm-Wave     & \multicolumn{2}{c}{500MHz} \\ \hline
			\textbf{Driving Scenario}                  & \multicolumn{1}{c|}{\textbf{Urban}} & \multicolumn{1}{c|}{\textbf{Rural}} & \textbf{Highway} \\ \hline
			\textbf{Dynamic Scenes}  & NA                & NA            & Scene 1-10          \\ \hline
			\textbf{Basestations}    & 4                 & 3             & 2         \\ \hline
			\textbf{Users}           & 63350               & 45250           &58610           \\ \hline
			\textbf{Size of Antenna} & \multicolumn{3}{c}{x=1; y=32; z=8}        \\ \hline
			\textbf{Antenna spacing} & \multicolumn{3}{c}{0.5$\lambda$}                 \\ \hline
			\textbf{Active Paths}    & \multicolumn{3}{c}{5}                         \\ \hline
			\textbf{Beams}           & \multicolumn{3}{c}{512}                       \\ \hline
		\end{tabular}%
}}%
\end{table}
\vspace{-10pt}

\subsection{Performance metrics and Selection of Loss functions}
To evaluate the effectiveness of the proposed CAR-BRAINet model, a comprehensive set of performance metrics is employed. Huber Loss (HL) is chosen as the primary loss function during training for its balance between Mean Squared Error (MSE) and Mean Absolute Error (MAE). HL is less sensitive to outliers than MSE while providing smoother gradients than MAE. \\
\indent The mathematical expressions of the loss metrics used in this study are given in Eq.(\ref{eqn:Eq(10)}), Eq.(\ref{eqn:Eq(11)}) and Eq.(\ref{eqn:Eq(12)}). Table~\ref{tab:my-table3} presents the loss values for each loss function, with the best results highlighted to justify the selection of Huber Loss.\\

\vspace{-0.2cm}
\begin{equation}
\mathbf{MAE} = \mathrm{\frac{1}{N}  \sum_{i=1}^N |y_{i}- \hat{y_{i}}|}
\label{eqn:Eq(10)}
\end{equation}
\begin{equation}
\mathbf{RMSE} = \mathrm{\sqrt{ \frac{1}{N}  \sum_{i=1}^N  (\hat{y_{i}}^{2} -  y_{i}^{2}) } }
\label{eqn:Eq(11)}
\end{equation}

\vspace{-0.2cm}
\begin{equation*}
\mathbf{x} = \mathrm{( y_{i} -  \hat{ y_{i} }  )}
\end{equation*}

\vspace{-0.2cm}
\begin{equation}
\mathbf{L_{ \delta } (x)}  =\begin{cases}\frac{1}{2} x^{2}&  for |x| \leq  \delta \\ \delta(|x|- \frac{1}{2} \delta)   & otherwise\end{cases}
\label{eqn:Eq(12)}
\end{equation}

where, \\
$y\textsubscript{i}$ = True label of the $i^{th}$ data sample \\
$\hat{ y_{i} }$ = Predicted label of the $i^{th}$ data sample\\
N = Total number of data samples\\

The value of $\delta$ is chosen to be 0.1 w.r.t context of this study.

\begin{table}[!h]
\normalsize
\caption{Selection of the loss function}
\label{tab:my-table3}
\resizebox{\columnwidth}{!}{%
	\renewcommand{\arraystretch}{1.5}{%
		\begin{tabular}{cccccc}
			\hline
			\multicolumn{1}{c|}{\multirow{2}{*}{\textbf{\begin{tabular}[c]{@{}c@{}}SL\\ No\end{tabular}}}} &
			\multicolumn{1}{c|}{\multirow{2}{*}{\backslashbox{\textbf{Loss}}{\textbf{Size of the Data}}}} &
			\multicolumn{2}{c|}{\textbf{50\%}} &
			\multicolumn{2}{c}{\textbf{90\%}} \\ \cline{3-6} 
			\multicolumn{1}{c}{} &
			\multicolumn{1}{c|}{} &
			\multicolumn{1}{c|}{\textbf{Sparse Users}} &
			\multicolumn{1}{c|}{\textbf{Dense Users}} &
			\textbf{Sparse Users} &
			\textbf{Dense Users} \\ \hline
			\cellcolor[HTML]{FDFF26}1. & \cellcolor[HTML]{FDFF26}\textbf{HL}   & \cellcolor[HTML]{FDFF26}$\mathbf{4.47 \times  10^{-4}}$    & \cellcolor[HTML]{FDFF26}$\mathbf{3.2\times10^{-5}}$    & \cellcolor[HTML]{FDFF26}$\mathbf{4.41\times10^{-4}}$ & \cellcolor[HTML]{FDFF26}$\mathbf{3.02\times10^{-5}}$   \\ \hline
			2. & \textbf{RMSE} & $7.24\times10^{-2}$    & $1.0552\times10^{-2}$ & $7.1\times10^{-2}$  & $1.0252\times10^{-2}$ \\ \hline
			3. & \textbf{MAE}  & $1.10227\times10^{-2}$ & $1.377\times10^{-3}$  & $1.14\times10^{-2}$ & $1.312\times10^{-3}$  \\ \hline
		\end{tabular}%
}}%
\end{table}

\vspace{-4pt}

\subsection{Experimental Setup}
The experiments are conducted using Keras~\cite{ref39} with TensorFlow~\cite{ref40} as the backend on 40 Core 2 x Xeon G-6248 processor with 2 x NVIDIA V100 Card. The proposed algorithm is trained with the hyper parameters listed in Table~\ref{tab:my-table4}. A learning rate scheduler, early stopping and L2 regulariser are employed to prevent over-fitting of the model.

\begin{table}[!h]
\caption{Training Hyper-parameters of CAR-BRAINet}
\label{tab:my-table4}
\centering
\tiny
\resizebox{0.75\columnwidth}{!}{%
		\begin{tabular}{l|cc}
			\hline
			\multicolumn{1}{c|}{\textbf{Parameters}}                                                                      & \multicolumn{2}{c}{\textbf{Value}}              \\ \hline
			\textbf{Optimiser}                                                                        & \multicolumn{2}{c}{ADAM}                        \\ \hline
			\multirow{3}{*}{\textbf{Learning rate}}        & \multicolumn{2}{c}{ReduceLRonPlateau}           \\ \cline{2-3} 
			& \multicolumn{1}{c|}{Patience}          & 10     \\ \cline{2-3} 
			& \multicolumn{1}{c|}{Min learning rate} & 0.0001 \\ \hline
			\textbf{Dropout}                                                                          & \multicolumn{2}{c}{0.2}                         \\ \hline
			\textbf{\begin{tabular}[l]{@{}c@{}}Multi-Head \\ Attention\end{tabular}}                                                                   & \multicolumn{1}{c|}{Heads}          & 64     \\ \hline
			\textbf{Regulariser}                                                                      & \multicolumn{1}{c|}{L2}                & 0.0001 \\ \hline
			\textbf{Epochs}                                                                           & \multicolumn{2}{c}{500}                         \\ \hline
			\textbf{Batch size}                                                                       & \multicolumn{2}{c}{100}                         \\ \hline
			\textbf{Data split}                                                                       & \multicolumn{2}{c}{80:20}                       \\ \hline
			\textbf{Loss}                                                     & \multicolumn{2}{c}{Huber Loss}                   \\ \hline
			\textbf{Early Stopping}                                                                   & \multicolumn{1}{c|}{Patience}          & 35     \\ \hline
		\end{tabular}%
	} 
	\end{table}
	\vspace{-3pt}

	\section{Simulation Results \& Discussion}\label{6}
	
	This section presents the experimental outcomes of the proposed CAR-BRAINet along with an exhaustive comparison of the developed model with existing state-of-the-art (SOTA) models, accompanied by relevant discussions.
	
	\vspace{-0.3cm}
	\subsection{Experimental Outcome}
	
\begin{figure}[!h]
\centering
\includegraphics[width=0.75\columnwidth,]{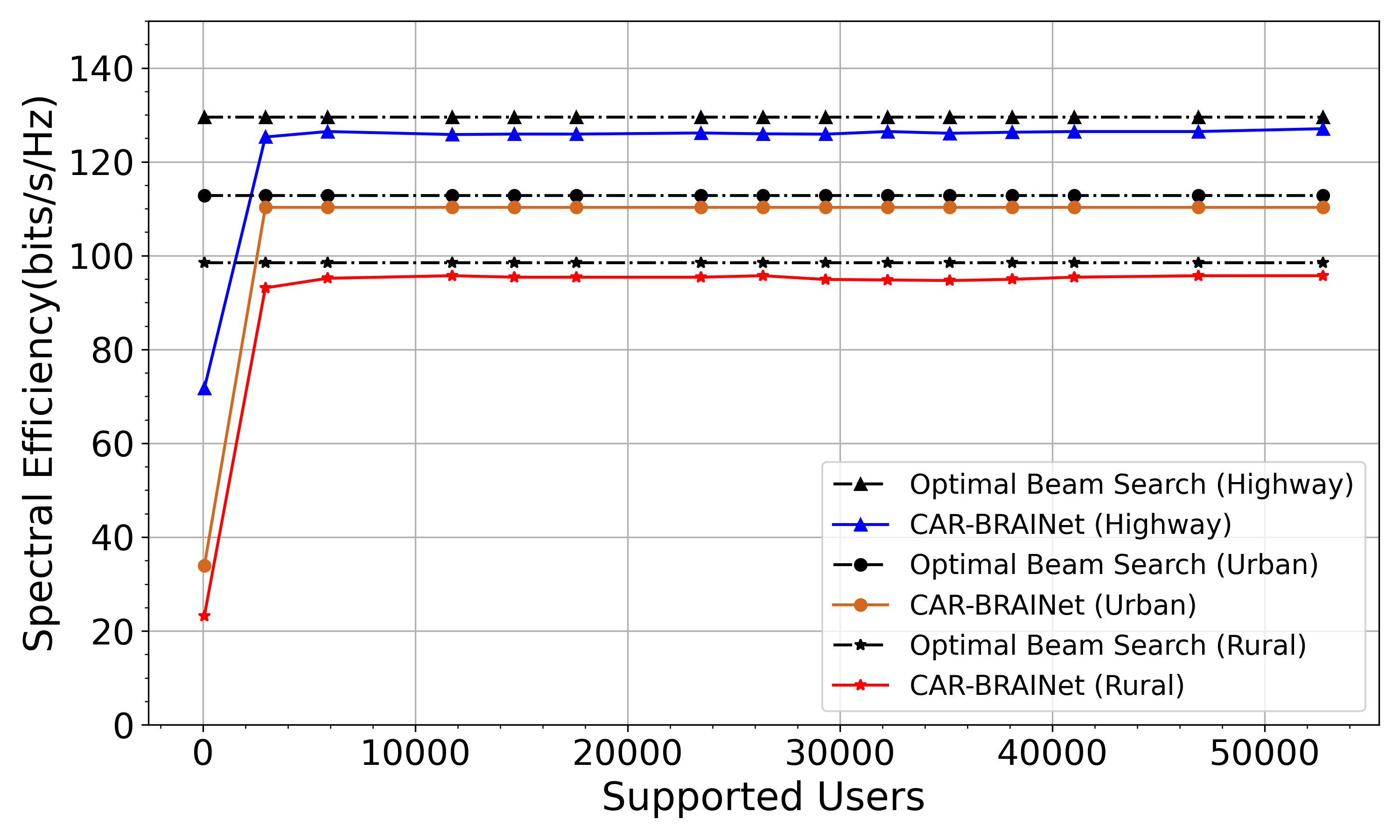}
\caption{Spectral Efficiency for C-V2X protocol}
\label{fig:Fig3}
\end{figure}

\begin{figure}[!h]
\centering
\includegraphics[width=0.75\columnwidth,]{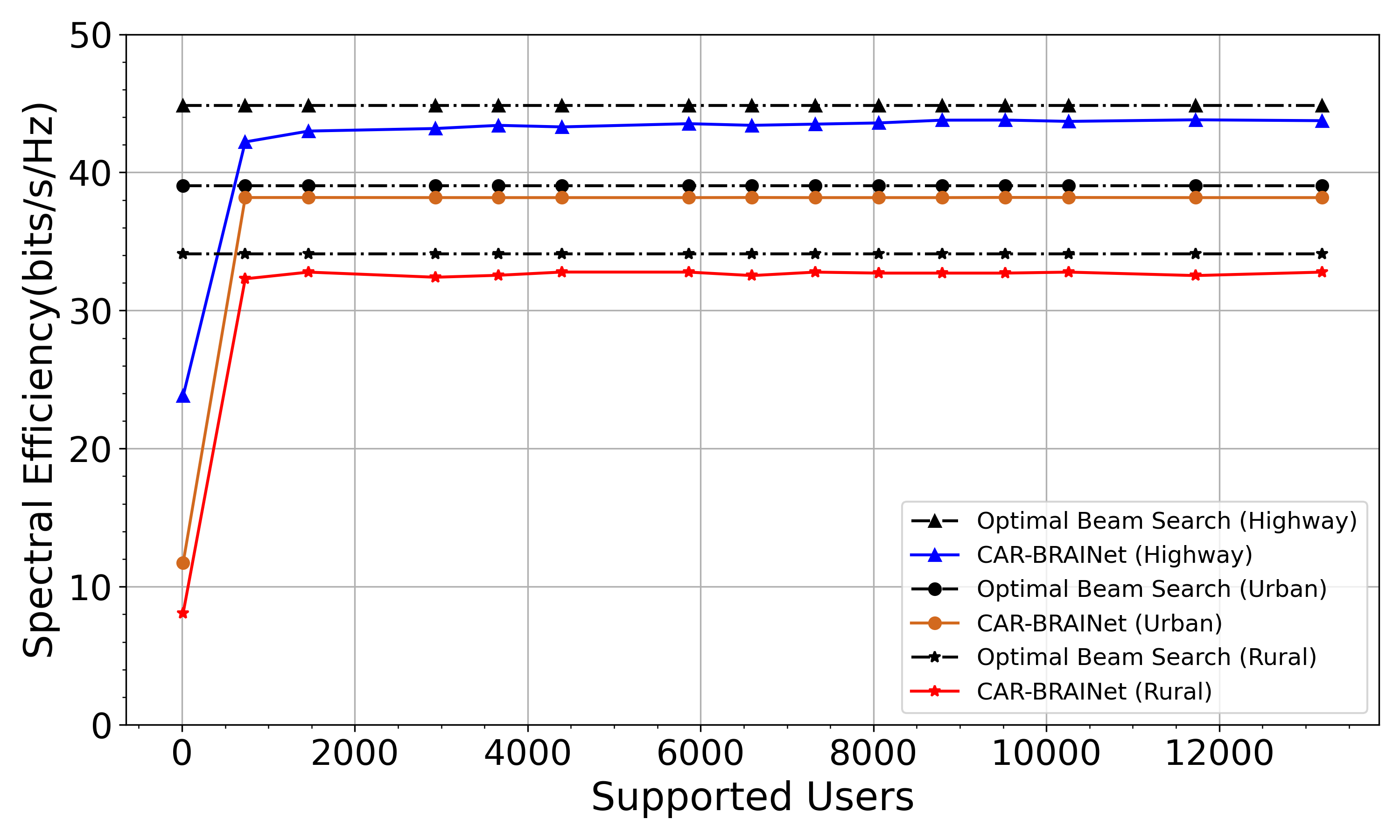}
\caption{Spectral Efficiency for IEEE 802.11BD protocol}  
\label{fig:Fig4}
\end{figure}

	As mentioned in Section \ref{3}, spectral efficiency is considered as the metric of interest to evaluate the effectiveness of the proposed CAR-BRAINet. Fig.~\ref{fig:Fig3} and Fig.~\ref{fig:Fig4} depicts the SE achieved for C-V2X and IEEE 802.11BD protocol respectively. A careful examination details the performance achieved by the proposed model under the three-diverse driving scenarios (URH). Even with the presence of harsh geographical blockages imposed by rural scenario, the CAR-BRAINet has presented a stable and impressive SE for both the MAC protocols considered. It is clear that CAR-BRAINet provides the highest SE in highway driving scenario, which proves the model's capacity to optimise the performance in the presence of dense traffic mostly made of highly mobile users. 
	
\begin{figure}[!h]
\centering
\includegraphics[width=0.75\columnwidth]{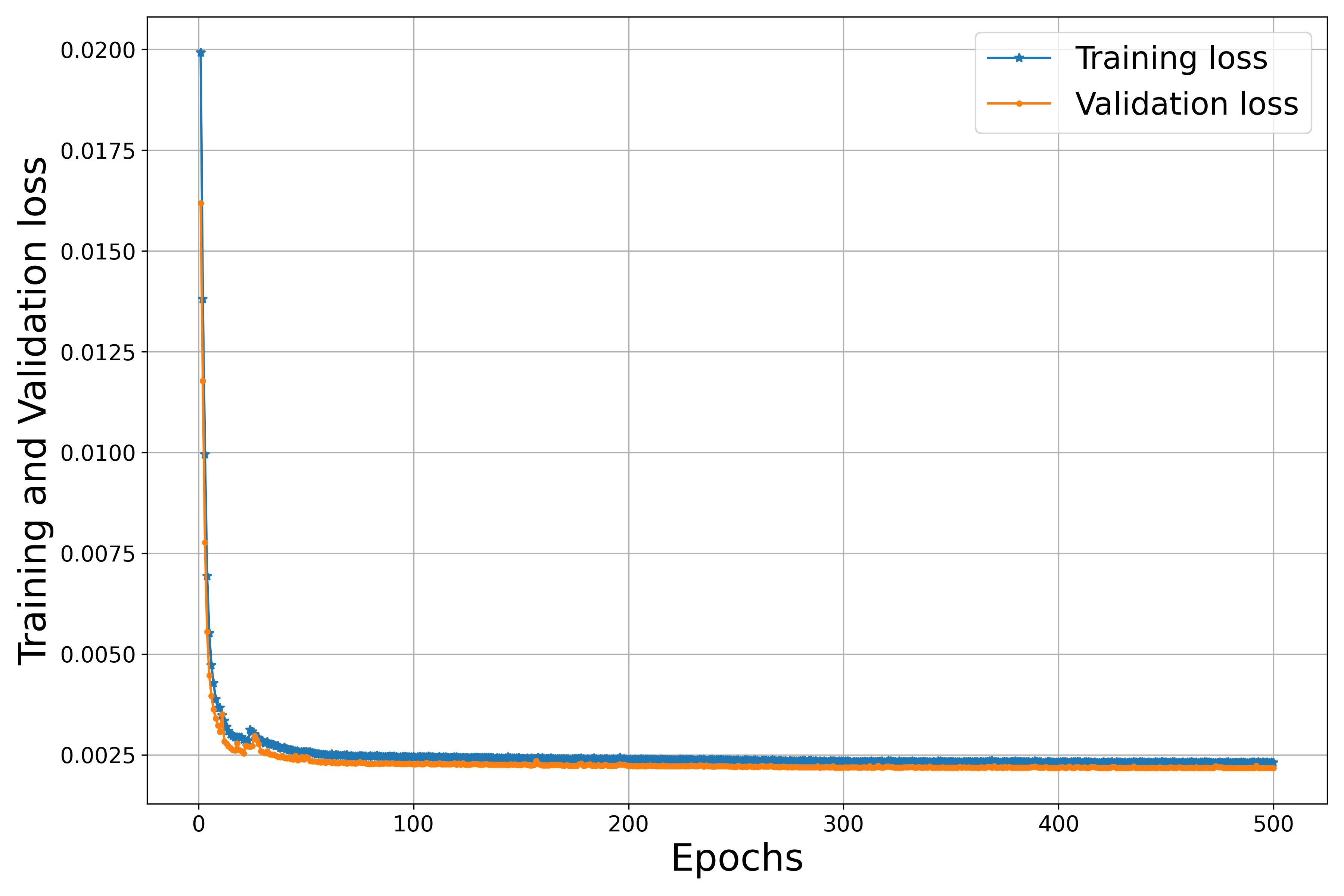}
\caption{Loss converged for the CAR-BRAINet}
\label{fig:Fig5}
\vspace{-2mm}
\end{figure}

As this study is based on regression analysis, achieving a validating loss curve, which adds to the stability and correctness of the proposed regression model, is essential. The training and validation loss recorded for this experimentation is shown in Fig.~\ref{fig:Fig5}. The steady decrease in loss is evident, affirming the model's effectiveness in delivering precise beam between the base-station and mobile users.

\begin{figure}[!h]
\centering
\includegraphics[width=0.75\columnwidth]{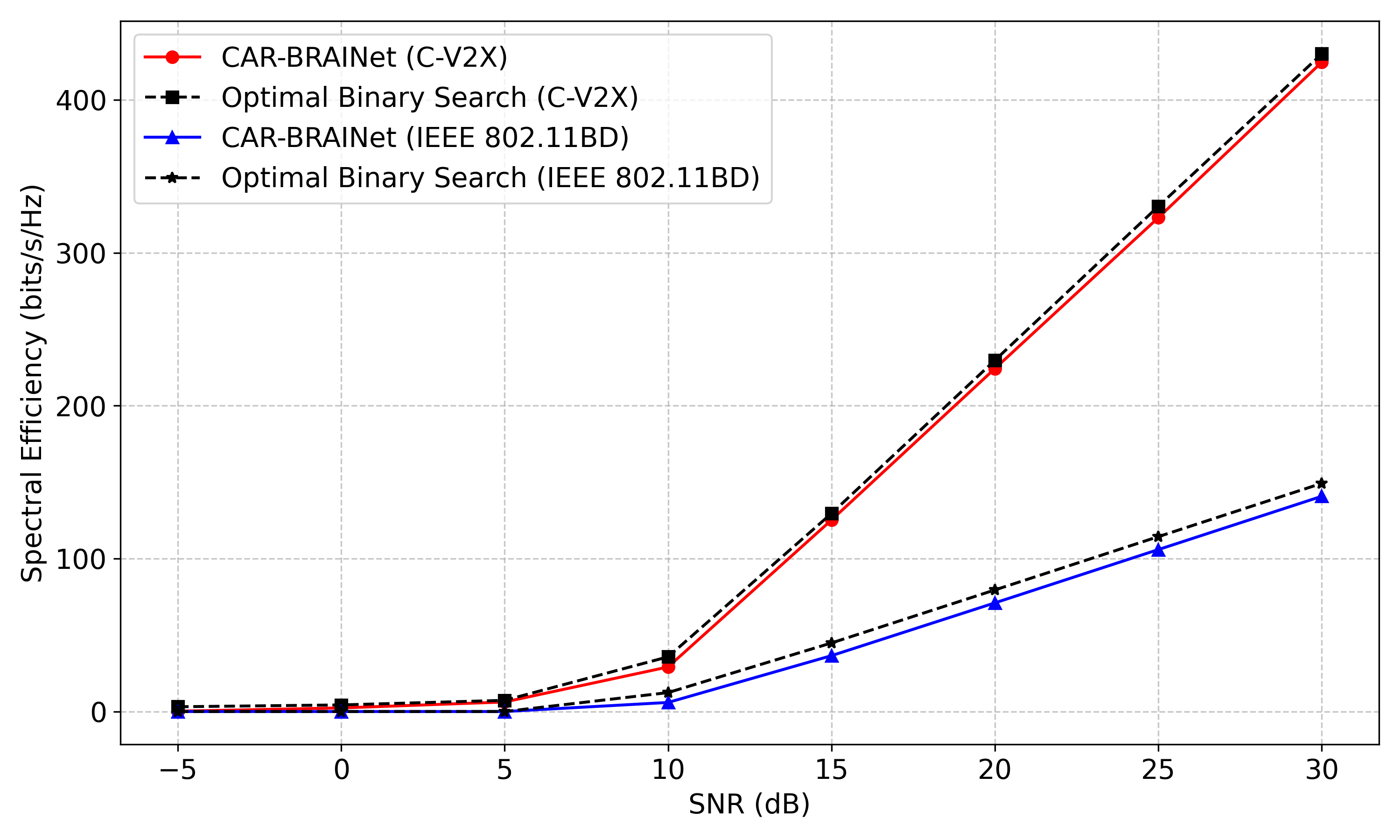}
\caption{Variation of Spectral Efficiency with SNR} 
\label{fig:Fig6}
\vspace{-2mm}
\end{figure}

Signal-to-Noise ratio (SNR) is one of the key indicator of signal quality in any communication system. The proposed beam prediction framework is tested under varying SNR values ranging from the least (-5 dB) to the possibly high value of 30 dB. Fig.~\ref{fig:Fig6} showcases the steady improvement in SE with the increasing SNR. This characteristic is studied under both the MAC protocols. 

\begin{figure}[!h]
\centering
\includegraphics[width=0.75\columnwidth]{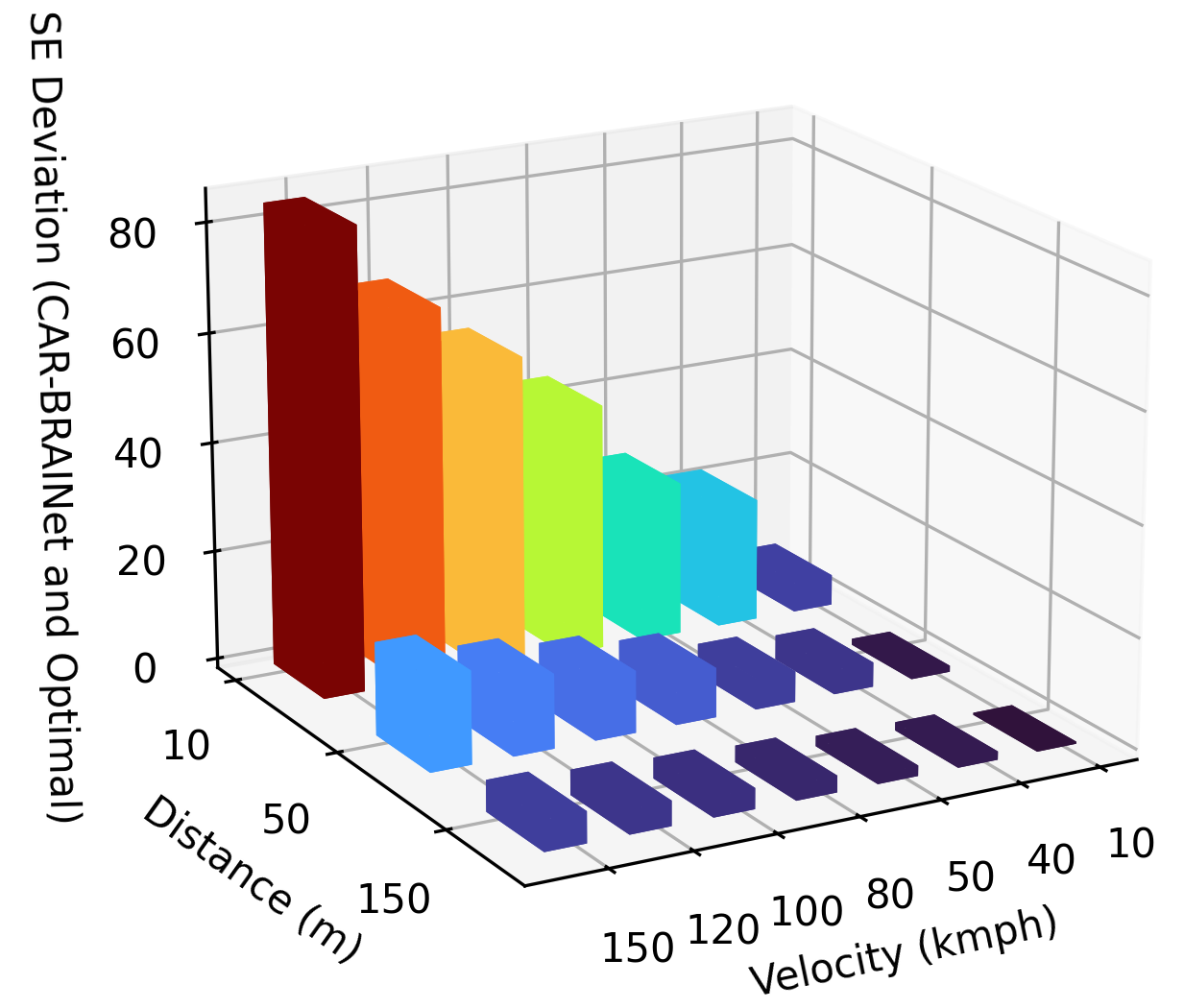}
\caption{Sensitivity of Spectral Efficiency w.r.t Distance and Velocity}  
\label{fig:Fig7}
\vspace{-2mm}
\end{figure}

As the proposed framework is to be deployed on an highly mobile environment, it is essential to study the performance of CAR-BRAINet on varying distance and velocity. A SE deviation between proposed and optimal beam prediction is plotted against varying distance and velocity in Fig.~\ref{fig:Fig7}. As the proposed model is based on the low-overhead design proposed in \cite{ref42_alkha}, it exhibits minimal dip for varying velocity and distance pairs. However, due to high overhead associated with the traditional beam prediction, the SE suffers higher dip for small variation in distance and velocity. Fig.~\ref{fig:Fig7} justifies the stability of the proposed CAR-BRAINet in stabilising a promising SE value even at a velocity of 150km/hr and for a distance of 150m, which is important to maintain a satisfactory quality of service (QoS) in a HetVNets.

\begin{figure}[!h]
	\centering
	\includegraphics[width=0.75\columnwidth]{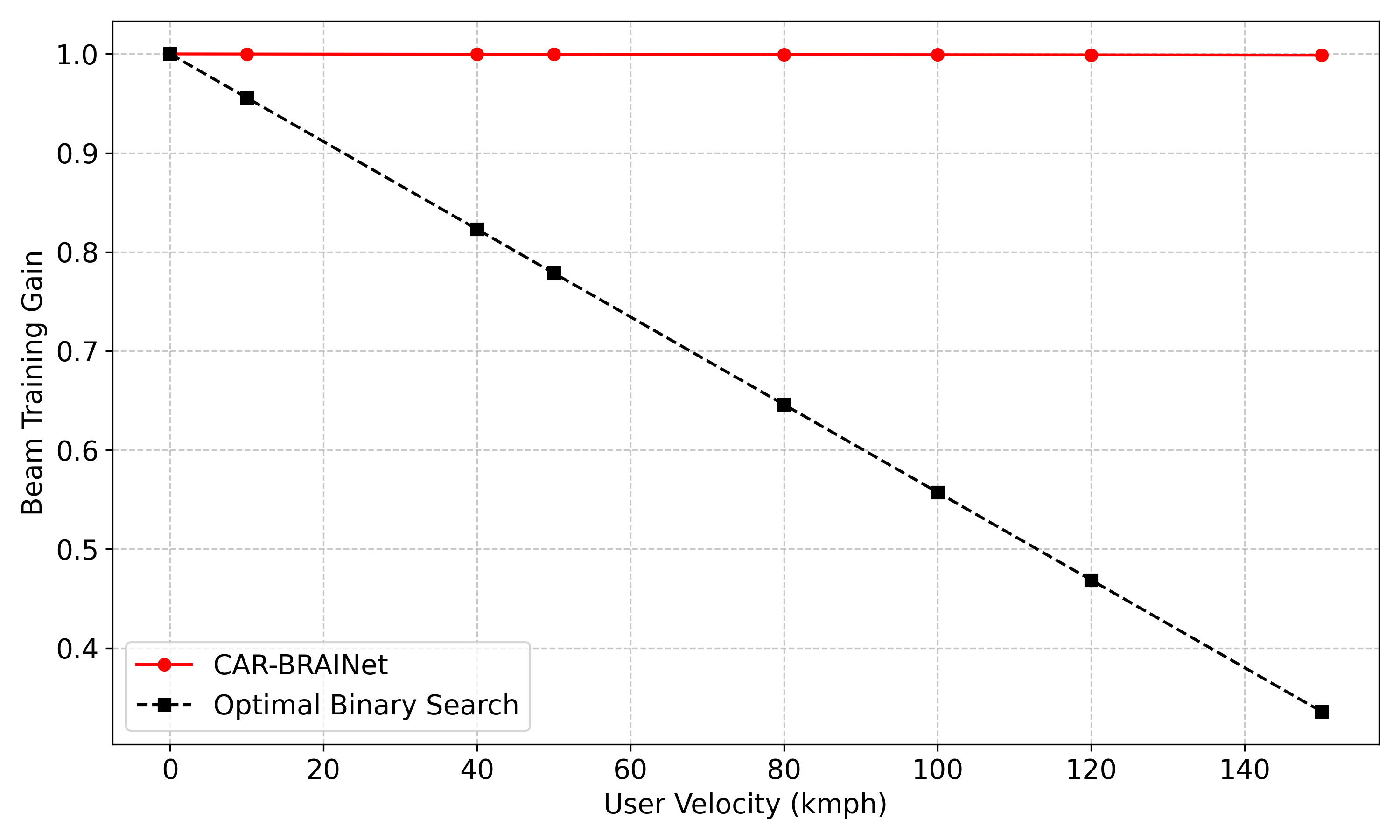}
	\caption{Variation of Beam training Gain with User velocity}
	\label{fig:Fig6a}
	\vspace{-2mm}
\end{figure}

Also, Figure.~\ref{fig:Fig6a} depicts the stability of beam training gain of the CAR-BRAINet against the traditional beam search technique with varying user velocity. 

\begin{figure}[!h]
	\centering
	\includegraphics[width=0.75\columnwidth]{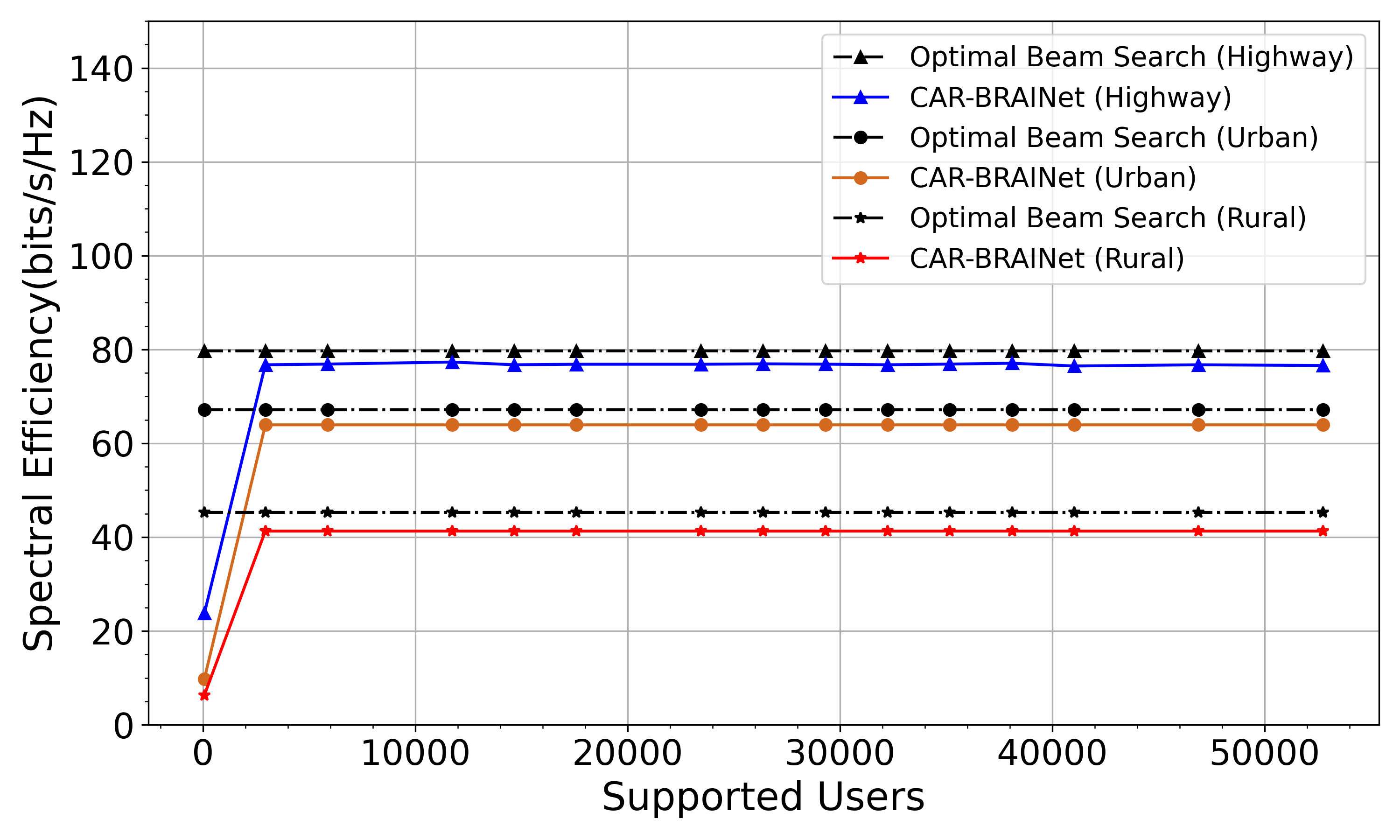}
	\caption{Impact of Doppler Effect on Spectral Efficiency}  
	\label{fig:Fig8}
	\vspace{-2mm}
\end{figure}

Doppler effect is associated with the high velocity mobile users. A steep degradation in performance is expected due to the presence of Doppler shift. Hence, it is crucial to examine the performance of CAR-BRAINet under an environment with Doppler effect. It is evident from Fig.~\ref{fig:Fig8} that CAR-BRAINet is successful in maintaining a satisfactorily good range of SE, which can sustain a quality link between the BST and MU. The effectiveness of the proposed model under the effect of Doppler shift is studied for the three diverse driving environments.

\subsection{Comparative Analysis}

In this section, a comparative analysis between the proposed CAR-BRAINet and leading SOTA\footnote{Few SOTA methods have not disclosed the respective values, due to which those parametric values are omitted from being mentioned on respective plots/tables} beam prediction methods. Critical dimensions such as performance, parameter count, training-testing time and converged loss are analysed to provide a fair comparison. The references are labelled as R1\cite{ref33}, R2\cite{ref49_IMP_sub6GHz_Federa}, R3\cite{ref37_Nearest_1_vry}, R4\cite{ref43_Imp_Meta}, R5\cite{ref10_1_VRY_IMP_PLS_REF}, R6\cite{ref32_near_imp_r12} and R7\cite{ref44_Imp_PARAMOUNT}.

\begin{figure}[!h]
	\centering
	\includegraphics[width=0.75\columnwidth]{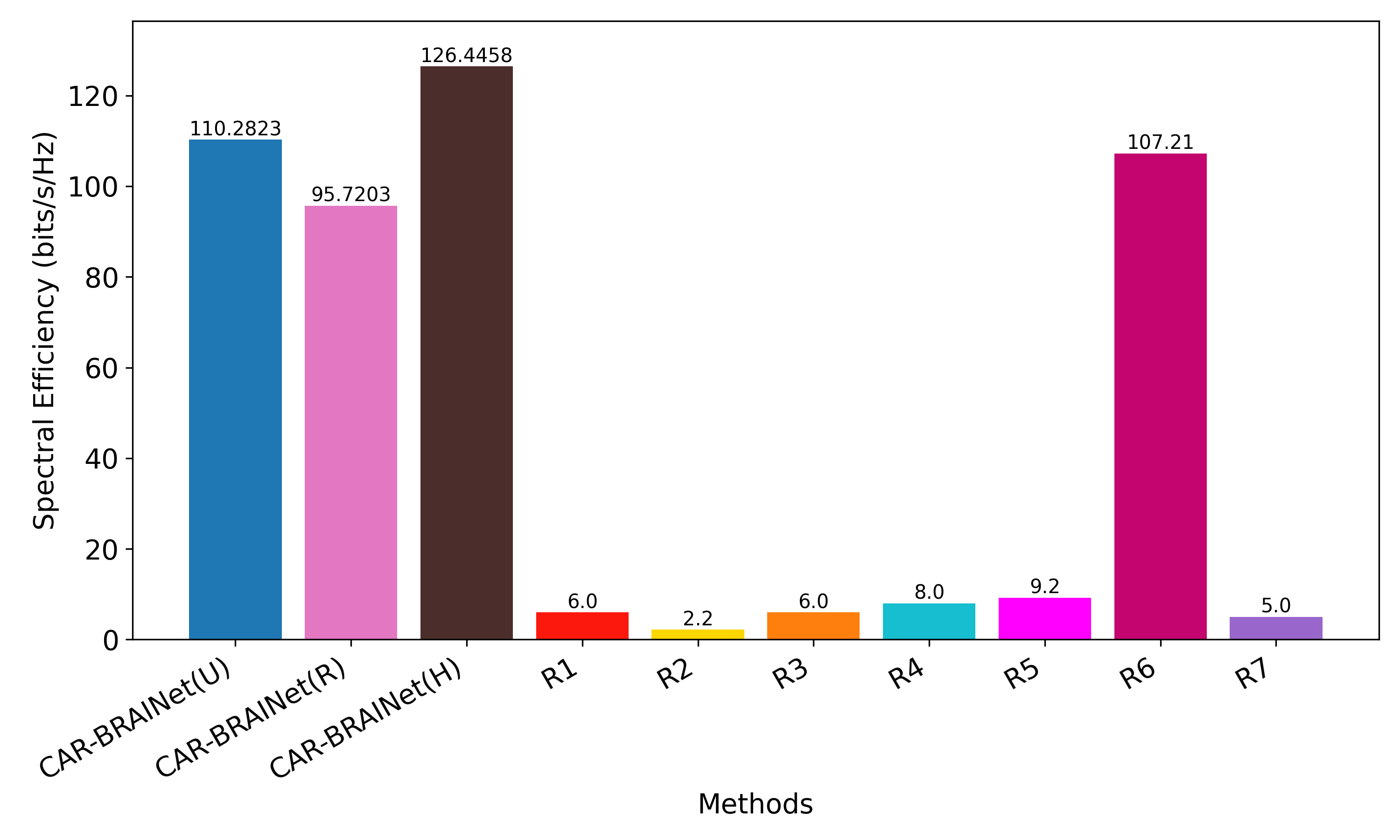}
	\caption{A comparative plot on Spectral efficiency}  
	\label{fig:Fig9}
	\vspace{-2mm}
\end{figure}

Fig.~\ref{fig:Fig9} records the spectral efficiency achieved by each of the models considered. It is clearly justifiable that the proposed model has put forth a stable and promising value of SE while outperforming the leading SOTA models by 17.9422\% of improvement. In order to avoid bias in the performance values, care is taken to compare SOTA models which share similar system parameters as that of this work.

\vspace{-0.1cm}

An improved performance at the cost of complex DL model is often overlooked due to its deployment complexity and latency. Table~\ref{tab:my-table5} provides an assessment of parameter count along with the loss and time factor associated with the proposed CAR-BRAINet over other beam prediction SOTA models considered. Noticeably, the highlighted row justifies the optimal trade off between the proposed model's complexity and performance. It is evident that CAR-BRAINet has showcased the least value of loss along with quick testing time, validating the model's stability and correctness in predicting the beam instantly with minimal errors. Thereby making it a suitable solution for beam prediction in diverse URH driving scenarios. Also, the proposed framework successfully defends the idea of developing a lightweight yet robust beam prediction algorithm.

\begin{table}[!h]
	\caption{Comparative Analysis on Evaluation Metrics}
	\label{tab:my-table5}
	\centering
	\resizebox{0.95\columnwidth}{!}{%
		\renewcommand{\arraystretch}{1.55}{%
			\begin{tabular}{ccccccc}
				\hline
				\multicolumn{1}{c|}{\multirow{2}{*}{\textbf{Method}}} & \multicolumn{1}{c|}{\multirow{2}{*}{\textbf{Model}}} &
				\multicolumn{1}{c|}{\multirow{2}{*}{\textbf{\begin{tabular}[c]{@{}c@{}}Parameters\\ (million)\end{tabular}}}} & \multicolumn{2}{c|}{\textbf{Time(s)}}      &
				\multicolumn{2}{c}{\textbf{Error}}  \\ 
				\cline{4-7} 
				\multicolumn{1}{c|}{}                                 & \multicolumn{1}{c|}{}                                &
				\multicolumn{1}{c|}{}                                &
				\multicolumn{1}{c|}{\textbf{Training}} & \textbf{Testing}    &
				\multicolumn{1}{|c|}{\textbf{MAE}} & \textbf{RMSE} \\ \hline
				
				R1    & MLP  &17.32    & 8910     & 48.91     &0.114  &0.0709 \\ \hline
				R2      & MLP  &8.78            & 5457     & 64.51     &0.0814  &0.0543 \\ \hline
				R3       & CNN-LSTM  &0.345       & 408608    & 241.45   &0.0764  &0.0412 \\ \hline
				R4 & MLP   &10.76        & 9798    & 67.26   &0.0967  &0.0682 \\ \hline
				R5       & CNN-BiLSTM   &1.65    & 370216    & 406.83     &0.0623  &0.0334 \\ \hline
				R6       & PGP   &0.370    & 2941    & 94.62     &0.01809 &0.0284 \\ \hline
				R7       & CNN  &0.340     & 2160    & 64.24     &0.0749 &0.0439 \\ \hline
				\cellcolor[HTML]{FDFF26}\textbf{CAR-BRAINet} & \cellcolor[HTML]{FDFF26}\textbf{CNN-MHA} &
				\cellcolor[HTML]{FDFF26}\textbf{0.289} & \cellcolor[HTML]{FDFF26}\textbf{1500} & \cellcolor[HTML]{FDFF26}\textbf{0.820357} &
				\cellcolor[HTML]{FDFF26}\textbf{0.001312}  &
				\cellcolor[HTML]{FDFF26}\textbf{0.010252} \\ \hline
				\multicolumn{7}{l}{ND = Not Disclosed; PGP = Parallel Gradient Projection}\\ \hline
			\end{tabular}%
	}}
\end{table}

A consolidated bench-marking analysis is recorded in Table~\ref{tab:my-table6} w.r.t the prominent parameters considered. Undeniably, the proposed model is able to deliver a promising performance while keeping an acceptable trade off between the model's complexity and robustness. To maintain a similar grounds of comparison, we have chosen SOTA models which are based on the idea of utilising sub-6GHz system alongside full/partial mm-wave technology.

\begin{table*}[!h]
	\caption{System parameter bench-marking with state-of-the-art methods}
	\label{tab:my-table6}
	\resizebox{2\columnwidth}{!}{%
		\renewcommand{\arraystretch}{1.45}{%
		\begin{tabular}{cccccccccccccc}
			\hline
			\multicolumn{1}{c|}{} &
			\multicolumn{1}{c|}{} &
			\multicolumn{3}{c|}{\textbf{Bandwidth (MHz)}} &
			\multicolumn{3}{c|}{} &
			\multicolumn{1}{c|}{} &
			\multicolumn{1}{c|}{} &
			\multicolumn{1}{c|}{} &
			\multicolumn{1}{c|}{} &
			\multicolumn{1}{c|}{} &
			\\ \cline{3-5}
			\multicolumn{1}{c|}{\multirow{-2}{*}{\textbf{Reference}}} &
			\multicolumn{1}{c|}{\multirow{-2}{*}{\textbf{Year}}} &
			\multicolumn{1}{c|}{\textbf{\begin{tabular}[c]{@{}c@{}}sub\\ 6GHz\end{tabular}}} &
			\multicolumn{1}{c|}{\textbf{\begin{tabular}[c]{@{}c@{}}mm\\ wave\end{tabular}}} &
			\multicolumn{1}{c|}{\textbf{DSRC}} &
			\multicolumn{3}{c|}{\multirow{-2}{*}{\textbf{SE(bits/s/Hz)}}} &
			\multicolumn{1}{c|}{\multirow{-2}{*}{\textbf{\begin{tabular}[c]{@{}c@{}}No. of \\ Beams\\\end{tabular}}}} &
			\multicolumn{1}{c|}{\multirow{-2}{*}{\textbf{HetVNet}}} &
			\multicolumn{1}{c|}{\multirow{-2}{*}{\textbf{\begin{tabular}[c]{@{}c@{}}Over\\ head\end{tabular}}}} &
			\multicolumn{1}{c|}{\multirow{-2}{*}{\textbf{\begin{tabular}[c]{@{}c@{}}Doppler\\ Shift\end{tabular}}}} &
			\multicolumn{1}{c|}{\multirow{-2}{*}{\textbf{\begin{tabular}[c]{@{}c@{}}Velocity\\ (kmph)\end{tabular}}}} &
			\multirow{-2}{*}{\textbf{\begin{tabular}[c]{@{}c@{}}Distance\\ (m)\end{tabular}}} \\ \hline
			R1 &
			2020 &
			20 &
			500 &
			\ding{55} &
			\multicolumn{3}{c}{6} &
			ND &
			\ding{55} &
			Low &
			\multicolumn{1}{c}{\ding{55}} &
			\ding{55} &
			\ding{55} \\ \hline
			R2 &
			2021 &
			20 &
			500 &
			\ding{55} &
			\multicolumn{3}{c}{2.2} &
			ND &
			\ding{55} &
			\ding{55} &
			\multicolumn{1}{c}{\ding{55}} &
			\ding{55} &
			\ding{55} \\ \hline
			R3 &
			2022 &
			\ding{55} &
			500 &
			\ding{55} &
			\multicolumn{3}{c}{6} &
			ND &
			\ding{55} &
			\ding{55} &
			\ding{55} &
			80 &
			\ding{55} \\ \hline
			R4 &
			2023 &
			20 &
			500 &
			\ding{55} &
			\multicolumn{3}{c}{8} &
			512 &
			\ding{55} &
			\ding{55} &
			\multicolumn{1}{c}{\ding{55}} &
			\ding{55} &
			\ding{55} \\ \hline
			R5 &
			2023 &
			\ding{55} &
			500 &
			\ding{55} &
			\multicolumn{3}{c}{9.2} &
			ND &
			\ding{51} &
			\ding{55} &
			\ding{55} &
			\ding{55} &
			\ding{55} \\ \hline
			R6 &
			2023 &
			500 &
			\ding{55} &
			\ding{55} &
			\multicolumn{3}{c}{107.21} &
			ND &
			\ding{51} &
			\ding{55} &
			\ding{55} &
			\ding{55} &
			\ding{55} \\ \hline
			R7 &
			2024 &
			80 &
			500 &
			\ding{55} &
			\multicolumn{3}{c}{5} &
			ND &
			\ding{55} &
			\ding{55} &
			\ding{55} &
			\ding{55} &
			\ding{55} \\ \hline

			\rowcolor[HTML]{FDFF26} 
			\cellcolor[HTML]{FDFF26} &
			\cellcolor[HTML]{FDFF26} &
			\cellcolor[HTML]{FDFF26} &
			\cellcolor[HTML]{FDFF26} &
			\multicolumn{1}{c|}{\cellcolor[HTML]{FDFF26}} &
			\multicolumn{1}{c|}{\cellcolor[HTML]{FDFF26}\textbf{U}} &
			\multicolumn{1}{c|}{\cellcolor[HTML]{FDFF26}\textbf{R}} &
			\multicolumn{1}{c|}{\cellcolor[HTML]{FDFF26}\textbf{H}} &
			\cellcolor[HTML]{FDFF26} &
			\cellcolor[HTML]{FDFF26} &
			\cellcolor[HTML]{FDFF26} &
			\cellcolor[HTML]{FDFF26} &
			\cellcolor[HTML]{FDFF26} &
			\cellcolor[HTML]{FDFF26} \\ \cline{6-8}
			\rowcolor[HTML]{FDFF26} 
			\multirow{-2}{*}{\cellcolor[HTML]{FDFF26}\textbf{CAR-BRAINet}} &
			\multirow{-2}{*}{\cellcolor[HTML]{FDFF26}\textbf{2025}} &
			\multirow{-2}{*}{\cellcolor[HTML]{FDFF26}\textbf{20}} &
			\multirow{-2}{*}{\cellcolor[HTML]{FDFF26}\textbf{500}} &
			\multicolumn{1}{c|}{\multirow{-2}{*}{\cellcolor[HTML]{FDFF26}\textbf{20}}} &
			\textbf{110.2823} &
			\textbf{95.7203} &
			\multicolumn{1}{c|}{\cellcolor[HTML]{FDFF26}\textbf{126.4458}} &
			\multirow{-2}{*}{\cellcolor[HTML]{FDFF26}\textbf{512}} &
			\multirow{-2}{*}{\cellcolor[HTML]{FDFF26}\ding{51}} &
			\multirow{-2}{*}{\cellcolor[HTML]{FDFF26}\textbf{Low}} &
			\multirow{-2}{*}{\cellcolor[HTML]{FDFF26}\ding{51}} &
			\multirow{-2}{*}{\cellcolor[HTML]{FDFF26}\textbf{150}} &
			\multirow{-2}{*}{\cellcolor[HTML]{FDFF26}\textbf{150}} \\ \hline
			\multicolumn{14}{l}{ND = Not Disclosed} \\ \hline
		\end{tabular}%
	}}%
\end{table*}

\section{Conclusion}\label{7}

This paper presents a lightweight DL-based beam prediction framework designed for Heterogeneous Vehicular Networks (HetVNets) of sub-6GHz, mm-wave and DSRC, addressing the challenges of maintaining reliable connectivity in high-mobility vehicular environments. The proposed method (CAR-BRAINet) integrates a novel feature extraction layer-multi-head attention (MHA), stacked with Convolutional Neural Network (CNN) to enhance spatial and temporal feature extraction. Since its challenging to extract noiseless CSI of mm-wave/DSRC channels, a combination of sub-6GHz and partial mm-wave/DSRC CSI is chosen as the feature-set for training CAR-BRAINet. The performance and effectiveness of the CAR-BRAINet is demonstrated with a aid of three-diverse high-quality non-stationary datasets pertaining to urban, rural and highway vehicular scenarios. These datasets are developed by accommodating prominent factors such as MAC layer protocols-(3GPP-C-V2X and IEEE 802.11BD), Doppler shifts, geographical blockages and highly varying velocity-distance profiles. The effectiveness of CAR-BRAINet in delivering a promising beam prediction mechanism, across sub-6GHz, mm-wave, and DSRC frequency bands is analysed by enhancing the spectral efficiency (SE) rendered by it. A steady improvement of 17.9422\% in SE with minimal beam training overhead and quick prediction time is observed in comparison to existing state-of-the-art techniques. Furthermore, the proposed method exhibits robustness under varying SNR levels, velocities, and distance for diverse vehicular scenarios, which proves model's scalability in adapting to heterogeneous use-cases. Also, as the CAR-BRAINet is independent of the location and angle subtended by the mobile users, redundant latency associated with these sensors are mitigated. These findings substantiates the potential of the proposed CAR-BRAINet as an efficient and benchmark beam prediction framework for next-generation 5G/B5G vehicular communication. Future work is focussed on extending the study to incorporate Reflective intelligent surface (RIS) equipped drones in HetVNets having Tera-Hertz communication system alongside sub-6GHz, mm-wave and DSRC.

\section{Acknowledgment}

This work is supported by the Centre for Development of Telematics (C-DOT), the Telecom R\&D Centre of the Department of Telecommunications (DoT), under the STAR Program (No. C-DOT/SP/04/2024). The authors also acknowledge the PARAM UTKARSH HPC facility of the National Supercomputing Mission, and thank the Centre for Cyber Physical Systems (CCPS), NIT Karnataka, Surathkal, for the financial support.

\bibliographystyle{IEEEtran}
\bibliography{bibiliography}

\section*{\textbf{Author Biography}}

\begin{wrapfigure}{l}{20mm} 
\includegraphics[width=1in,height=1in,clip,keepaspectratio]{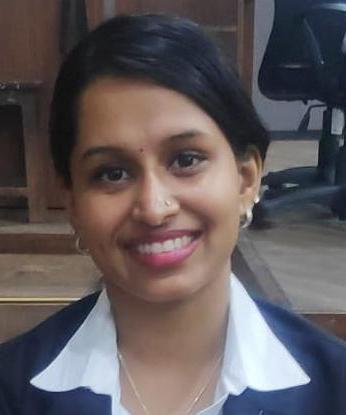}
\end{wrapfigure}\par
\fontsize{8}{10}\selectfont{\textbf{Aathira G Menon} is currently pursuing her Doctoral studies in the Department of Electronics and Communication Engineering at the National Institute of Technology Karnataka, Surathkal. She earned her M.Tech degree in Power Electronics \& Drives from Manipal Institute of Technology, Manipal, Karnataka, in 2021, and her B.E. in Electronics \& Communication Engineering from NMAM Institute of Technology, Nitte, Karnataka, in 2019.  Her current area of research lies in the design of Deep learning architectures for Vehicular communication, Wireless communication and Smart Cities.\\ \par

\begin{wrapfigure}{l}{15mm} 
\includegraphics[width=1in,height=1in,clip,keepaspectratio]{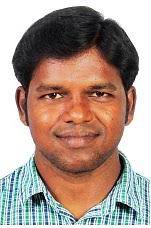}
\end{wrapfigure}
\textbf{Prabu Krishnan} (Senior Member, IEEE) is an Associate Professor in the Department of Electronics and Communication Engineering at the National Institute of Technology Karnataka (NIT-K) in Surathkal. Prior to this, he was an Associate Professor at VIT University Vellore’s SENSE, EEC, SRM Group, LICET, the Loyola group of institutions, and SGC Services Pvt. Ltd. in the CIPA Project NIC Puducherry. Overall he has 11+ years of experience with the wireless and optical communications domain. He is a notable alumnus of Anna University, where he graduated and post graduated. He pursued his PhD at NIT Trichy. Prof. Prabu mentored around 40 PhD and Masters students. He disseminated pertinent knowledge through 81 technical papers, 66 international journals, 15 international conferences, 2 book chapters, and a book, 100+ invited or conference talks. He received fellowships from the University Grant Commission (UGC) and the Technical Education Quality Improvement Programme (TEQIP). He is one among the top 2\% of scientists worldwide, as acknowledged by Elsevier and Stanford University US consecutively in the years 2019, 2020, 2021, and 2022. His research interests include Wireless Optical Communication (FSO, VLC, and Underwater), Optical Sensors, Nano-Photonics, 5G, Antennas and 6G–IoT. \\ \par

\begin{wrapfigure}{l}{20mm} 
\includegraphics[width=1in,height=1in,clip,keepaspectratio]{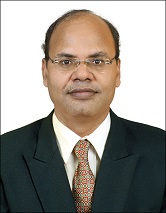}
\end{wrapfigure}
\textbf{Shyam Lal} received the M.Tech degree in Electronics and Communication Engineering from NIT Kurukshetra, Kurukshetra, India, in 2007, and the Ph.D. degree in image processing from BIT Mesra, Ranchi, India, in 2013. He has been working as an Associate Professor in the Department of Electronics \& Communication Engineering, National Institute of Technology Karnataka, Mangalore, India. His research interests include Artificial Intelligence, Machine and Deep Learning Algorithms for Satellite Data Processing and Analysis, Synthetic Aperture Radar (SAR) Data Processing and Analysis, Medical Image Processing applications, Cyber Security and Internet of things (IoT) applications, 5G \& B5G Wireless Sensor Networks applications processing.\par

\end{document}